\documentclass[english,useAMS,usenatbib]{mn2e} 
\usepackage{lmodern}

\usepackage[T1]{fontenc}
\usepackage[utf8]{inputenc}
\usepackage{color}
\usepackage[usenames,dvipsnames]{xcolor}
\usepackage{amsmath}
\usepackage{amssymb}
\usepackage{graphicx}
\usepackage{esint}
\usepackage{placeins}
\usepackage[authoryear]{natbib}

\usepackage{hyperref}
\definecolor{darkgreen}{rgb}{0.1,0.6,0.7}
\hypersetup{colorlinks = true,urlcolor=blue,citecolor=Blue}

\makeatletter
\usepackage{lmodern}\usepackage{epstopdf}

\let\jnl@style=\rm
\def\ref@jnl#1{{\jnl@style#1}}

\def\aj{\ref@jnl{AJ}}                   
\def\actaa{\ref@jnl{Acta Astron.}}      
\def\araa{\ref@jnl{ARA\&A}}             
\def\apj{\ref@jnl{ApJ}}                 
\def\apjl{\ref@jnl{ApJ}}                
\def\apjs{\ref@jnl{ApJS}}               
\def\ao{\ref@jnl{Appl.~Opt.}}           
\def\apss{\ref@jnl{Ap\&SS}}             
\def\aap{\ref@jnl{A\&A}}                
\def\aapr{\ref@jnl{A\&A~Rev.}}          
\def\aaps{\ref@jnl{A\&AS}}              
\def\azh{\ref@jnl{AZh}}                 
\def\baas{\ref@jnl{BAAS}}               
\def\bac{\ref@jnl{Bull. astr. Inst. Czechosl.}}
\def\caa{\ref@jnl{Chinese Astron. Astrophys.}}
\def\cjaa{\ref@jnl{Chinese J. Astron. Astrophys.}}
\def\icarus{\ref@jnl{Icarus}}           
\def\jcap{\ref@jnl{J. Cosmology Astropart. Phys.}}
\def\jrasc{\ref@jnl{JRASC}}             
\def\memras{\ref@jnl{MmRAS}}            
\def\mnras{\ref@jnl{MNRAS}}             
\def\na{\ref@jnl{New A}}                
\def\nar{\ref@jnl{New A Rev.}}          
\def\pra{\ref@jnl{Phys.~Rev.~A}}        
\def\prb{\ref@jnl{Phys.~Rev.~B}}        
\def\prc{\ref@jnl{Phys.~Rev.~C}}        
\def\prd{\ref@jnl{Phys.~Rev.~D}}        
\def\pre{\ref@jnl{Phys.~Rev.~E}}        
\def\prl{\ref@jnl{Phys.~Rev.~Lett.}}    
\def\pasa{\ref@jnl{PASA}}               
\def\pasp{\ref@jnl{PASP}}               
\def\pasj{\ref@jnl{PASJ}}               
\def\rmxaa{\ref@jnl{Rev. Mexicana Astron. Astrofis.}}%
\def\qjras{\ref@jnl{QJRAS}}             
\def\skytel{\ref@jnl{S\&T}}             
\def\solphys{\ref@jnl{Sol.~Phys.}}      
\def\sovast{\ref@jnl{Soviet~Ast.}}      
\def\ssr{\ref@jnl{Space~Sci.~Rev.}}     
\def\zap{\ref@jnl{ZAp}}                 
\def\nat{\ref@jnl{Nature}}              
\def\iaucirc{\ref@jnl{IAU~Circ.}}       
\def\aplett{\ref@jnl{Astrophys.~Lett.}} 
\def\apspr{\ref@jnl{Astrophys.~Space~Phys.~Res.}}
\def\bain{\ref@jnl{Bull.~Astron.~Inst.~Netherlands}} 
\def\fcp{\ref@jnl{Fund.~Cosmic~Phys.}}  
\def\gca{\ref@jnl{Geochim.~Cosmochim.~Acta}}   
\def\grl{\ref@jnl{Geophys.~Res.~Lett.}} 
\def\jcp{\ref@jnl{J.~Chem.~Phys.}}      
\def\jgr{\ref@jnl{J.~Geophys.~Res.}}    
\def\jqsrt{\ref@jnl{J.~Quant.~Spec.~Radiat.~Transf.}}
\def\memsai{\ref@jnl{Mem.~Soc.~Astron.~Italiana}}
\def\nphysa{\ref@jnl{Nucl.~Phys.~A}}   
\def\physrep{\ref@jnl{Phys.~Rep.}}   
\def\physscr{\ref@jnl{Phys.~Scr}}   
\def\planss{\ref@jnl{Planet.~Space~Sci.}}   
\def\procspie{\ref@jnl{Proc.~SPIE}}   

\newcommand{\half}{\frac{1}{2}}

\newcommand{\bomu}{\ensuremath{\boldsymbol{\mu}}}
\newcommand{\bou}{\ensuremath{\boldsymbol{u}}}

\newcommand{\boSig}{\ensuremath{{\sf{\Sigma}}}}

\newcommand{\sfA}{\ensuremath{{\sf{A}}}}

\newcommand{\both}{\ensuremath{\boldsymbol{\theta}}}

\newcommand{\calP}{\ensuremath{\mathcal{P}}}

\newcommand{\fatx}{\ensuremath{\boldsymbol{x}}}

\newcommand{\sfB}{\ensuremath{{\sf{B}}}}
\newcommand{\sfL}{\ensuremath{{\sf{L}}}}
\newcommand{\sfR}{\ensuremath{{\sf{R}}}}

\newcommand{\brsfL}{\ensuremath{{\breve{\sf{L}}}}}

\newcommand{\sfQ}{\ensuremath{{\sf{Q}}}}
\newcommand{\sfLam}{\ensuremath{{\sf{\Lambda}}}}

\newcommand{\sfC}{\ensuremath{{\sf{C}}}}
\newcommand{\invsfC}{\ensuremath{{\sf{C}^{-1}}}}

\newcommand{\brinvsfC}{\ensuremath{{\breve{\sf{C}}^{-1}}}}

\newcommand{\sfX}{\ensuremath{{\sf{X}}}}

\title[Blinding in astronomical data analysis]{A blinding solution for inference from astronomical data}

\author[E. Sellentin]{Elena Sellentin\\
Leiden Observatory, Leiden University, Huygens Laboratory, Niels Bohrweg 2, NL-2333 CA Leiden, The Netherlands.\\
}

\makeatother

\usepackage{babel}
\begin{document}
\setlength{\voffset}{-12mm} 

\date{Accepted: -$\infty$. Received: $-2\infty$; in original form: $t_0$.}

\maketitle
\pagerange{\pageref{firstpage}--\pageref{lastpage}} \pubyear{2016}

\label{firstpage} 
\begin{abstract}
This paper presents a joint blinding and deblinding strategy for inference of physical laws from astronomical data. The strategy allows for up to three blinding stages, where the data may be blinded, the computations of theoretical physics may be blinded, and --assuming Gaussianly distributed data-- the covariance matrix may be blinded. We found covariance blinding to be particularly effective, as it enables the blinder to determine close to exactly where the blinded posterior will peak. Accordingly, we present an algorithm which induces posterior shifts in predetermined directions by hiding untraceable biases in a covariance matrix. The associated deblinding takes the form of a numerically lightweight post-processing step, where the blinded posterior is multiplied with deblinding weights. We illustrate the blinding strategy for cosmic shear from KiDS-450, and show that even though there is no direct evidence of the KiDS-450 covariance matrix being biased, the famous cosmic shear tension with Planck could easily be induced by a mischaracterization of correlations between $\xi_-$ at the highest redshift and all lower redshifts. The blinding algorithm illustrates the increasing importance of accurate uncertainty assessment  in astronomical inferences, as otherwise involuntary blinding through biases occurs. 
\end{abstract}


\section{Introduction}
Astronomy provides many data sets that enable us to infer physical laws on energy-, size-, and time scales that are inaccessible to Earth-bound laboratories. Of particular interest to inference are astronomical observations which are often so rare that only a few -- or even no -- comparable observations are expected in a human lifetime. 
Amongst such unique data sets rank e.g.~observations of our Milky Way \citep{GaiaDR1,GaiaDR2} or across the entire cosmos \citep{Planck2018,Synergies,Euclid}, where by definition no second data set like the original will ever exist. 

Many astronomers therefore remember e.g.~peculiar stars in the Milky Way, the cold spot in the cosmic microwave background, or other directly visible features in the data, such as the `Great Wall' \citep{GreatWall} in galaxy surveys. Inference, on the other hand, is the attempt to go beyond the directly visible, and assigns a credibility to a hidden, unobservable quantity, such as a model of theoretical physics, or the values of free physical parameters.

Due to remembering data features on the one hand, but also due to iterative data cleaning to handle unexpected systematics, the physics inferred from astronomical data is sometimes regarded with scepticism. Post-dictions and retrospectively adapted models rank amongst frequently encountered points of critique. Iterative and often subconscious tampering with the analysis constitute further elements of concern, as they might lead to the confirmation of prior held beliefs \citep{Croft,Seehars}. The wish to avoid that such biases impact the inferred physics is therefore becoming widespread, and can be addressed by conducting analyses blindly.

Blinding strategies extend an analysis such that it becomes impossible to predict which physics will be discovered from it. Blinding can be extremely difficult to achieve. Indeed, most blinding techniques operate exclusively on the data, and therefore either interfere with the inevitable necessity to make informed decisions when cleaning astronomical data --- or are easily spotted as counterfeits.

Thus, the aim of this paper is to establish a blinding technique which meets the needs of astronomical inference. The raw data are left untouched, but the transition to science-ready data may be blinded if this does not cause substantial costs. Additionally, the computations of physical models may also be blinded. Crucially though, we find blinding of a likelihood by biasing a covariance matrix provides a very powerful third tier. It enables the blinder to specify nearly perfectly where the blinded posterior peaks, whilst causing negligible numerical costs.

In Sect.~\ref{sec::special}, we detail why astronomical inference both requires and enables special blinding strategies. The up to three-stage blinding algorithm is described in Sect.~\ref{sec::three}. In Sect.~\ref{sec::cov}, Sect.~\ref{sec::blindalg} and Sect.~\ref{sec::disable} we develop the algorithm for blinding a covariance matrix. The associated deblinding  is described in Sect.~\ref{sec::blindpost} and Sect.~\ref{sec::deblinding}. Throughout the paper, we demonstrate the algorithm on the data of \citet{KiDS450}, which is a cosmic shear analysis of the Kilo-Degree Survey (KiDS) further described in \citet{Kuijken2015, FenechConti}. The KiDS data\footnote{The public data products are available at http://kids.strw.leidenuniv.nl/cosmicshear2016.php} are processed by \textsc{THELI} \citep{Theli} and Astro-\textsc{WISE} \citep{Begeman,dejong}. Shears were measured with \textsc{lensfit} \citep{LensFit}, and photometric redshifts were obtained from PSF-matched photometry and calibrated using external overlapping spectroscopic surveys (see \citet{KiDS450}). The essentials of cosmic shear are summarized in appendix \ref{app:shear}. Appendix \ref{Kids_specific} summarizes the implications of this paper's findings for the KiDS-450 survey and its reported mild tension with Planck \citep{Planck2018}.

\section{Why astronomy has special blinding needs}
\label{sec::special}
We briefly review blinding techniques used in neighbouring fields, and contrast them with astronomy.

Popular blinding techniques in particle physics include masking of a `signal region', as e.g.~carried out by CMS  and ATLAS during the discovery of the Higgs boson \citep{HiggsCMS,HiggsATLAS}. Also the ANTARES neutrino telescope blinded a spatial signal region when studying excessive neutrino flux from the galactic ridge \citep{Antares2016}. The Large Underground Xenon experiment LUX \citep{LUX} and the gravitational wave facility LIGO \citep{LIGO} instead injected artificial signals into the detector, an approach known as `salting'.

\begin{figure*}
\includegraphics[width=0.89\textwidth]{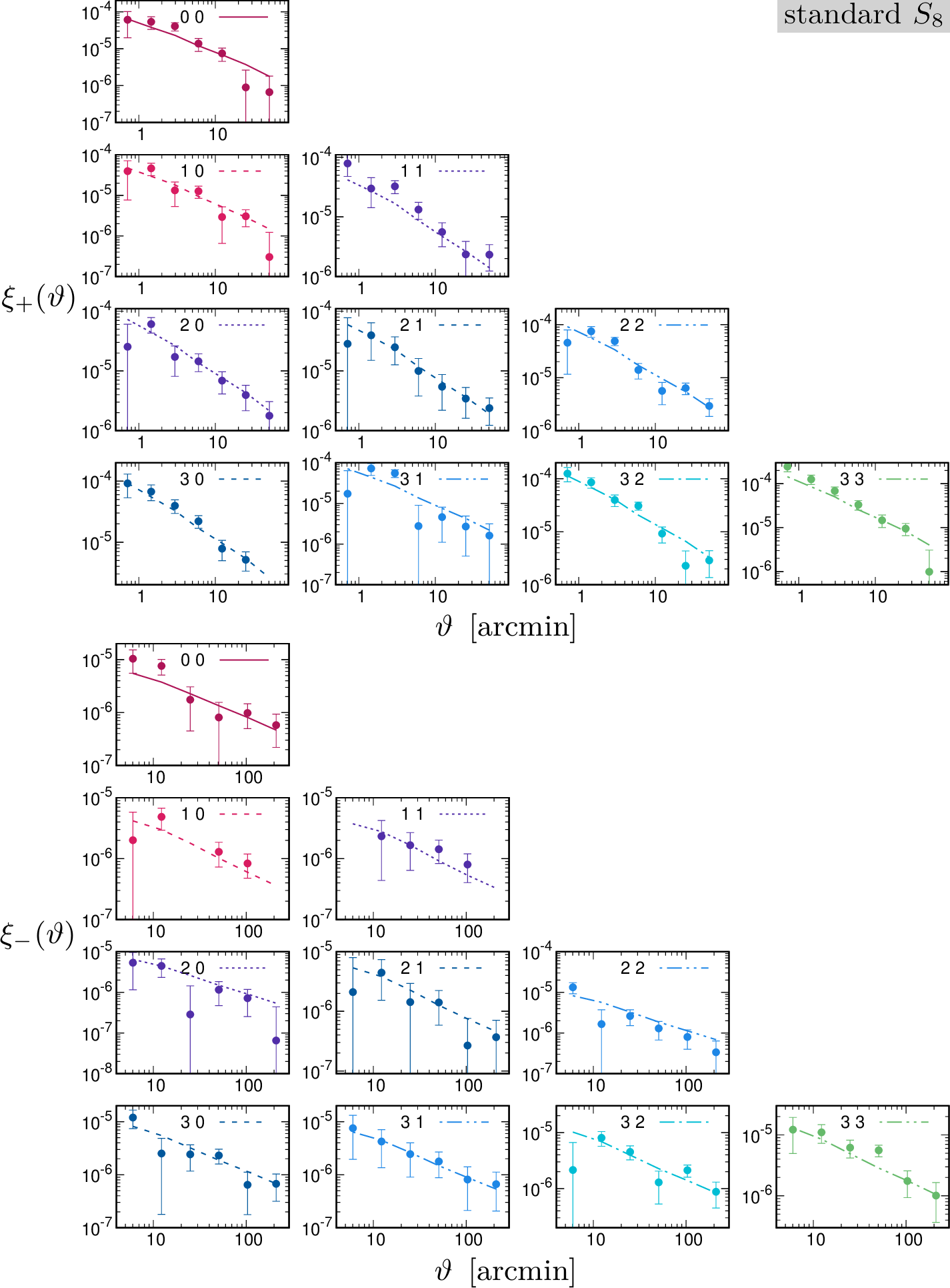} 
\caption{Example of an unblinded analysis: Joint plot of the original KiDS-450 data \citep{KiDS450}, the original error bars, and the original best-fitting theory curve, computed with $\Omega_{\rm m} = 0.2$ and $\sigma_8 = 0.838$. The upper triangle depicts the cosmic shear correlation function $\xi_+(\vartheta)$ over all redshift bin combinations as labelled; the lower triangle depicts the correlation function $\xi_-(\vartheta)$.}
\label{UnblindedBestFit}
\end{figure*}

\begin{figure*}
\includegraphics[width=0.89\textwidth]{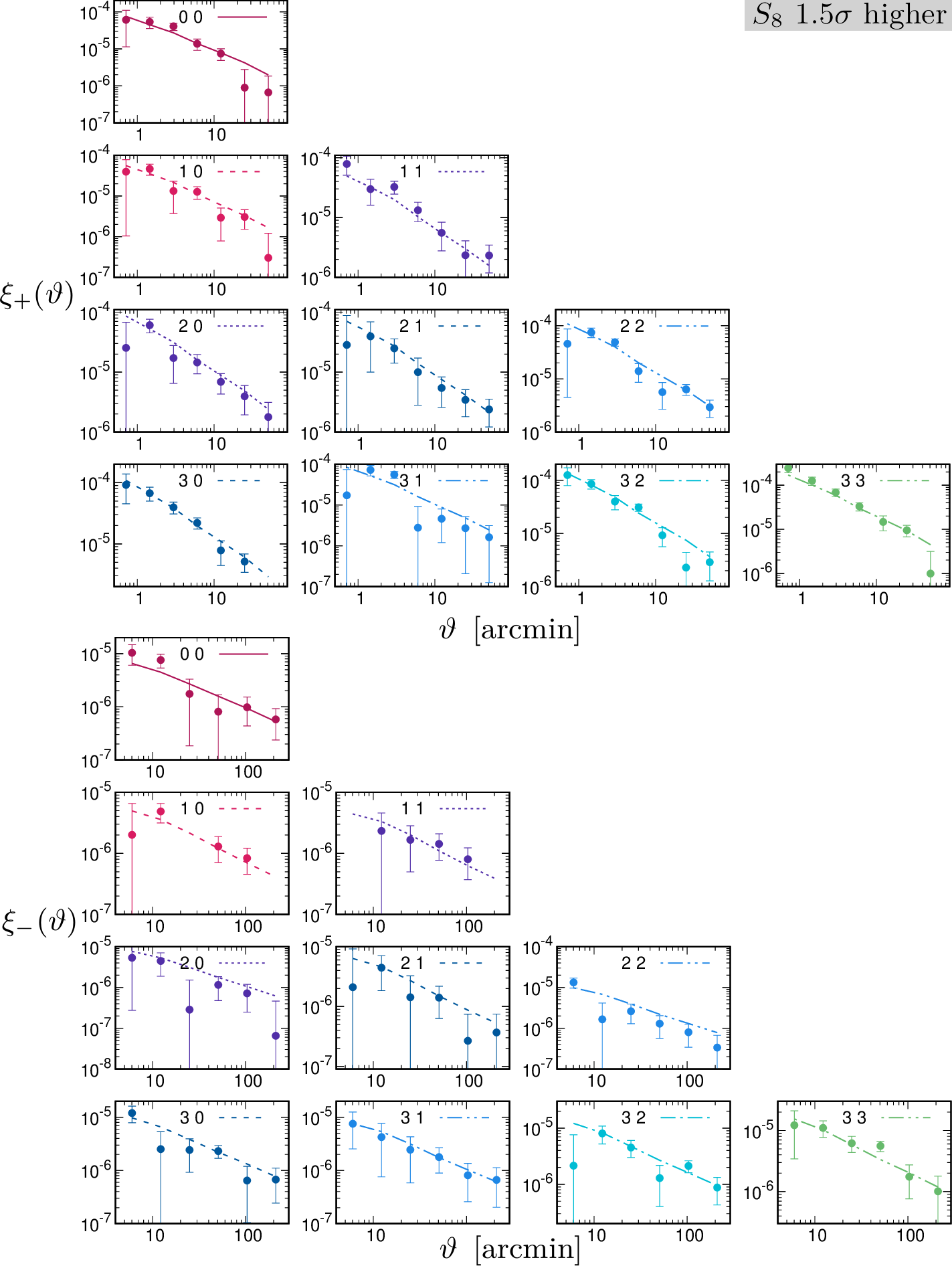} 
\caption{Example of a blinded analysis: The data are the same as in Fig.~\ref{UnblindedBestFit}, but the error bars differ, as the analysis was blinded by biasing the covariance matrix to enforce a shift of the posterior by about 1.5 posterior standard deviations in $S_8$. The now plotted theory curve is the peak of the blinded posterior, which lies at $\Omega_{\rm m } = 0.2748, \sigma_8 =  0.7548$. Comparison with Fig.~\ref{UnblindedBestFit} illustrates that `fitting by eye' is impossible and that it is extremely difficult to tell the blinded and unblinded analysis apart.}
\label{HolyBlindBestFit}
\end{figure*}

Common to these blinding techniques is their direct operation on the raw data, either by masking or by imitating signals. This is effective when new physics leaves visible imprints in a particle physics experiment: Novel particles may cause unconventional tracks in a detector, and decaying new particles will cause visible peaks above the detector's background, called `resonances'. Salting and blinding of a signal region effectively masks the presence of such features.

In contrast, the majority of astronomical data sets do not exhibit a split into a signal- and a background region, thereby having nothing whose hiding would be of any advantage.  The lack of a signal region is e.g.~illustrated by  Fig.~\ref{UnblindedBestFit}, which depicts the cosmic shear data set from \citet{KiDS450}, from which the best-fitting parameters for the cosmic dark matter density $\Omega_{\rm m}$ was determined to be $\Omega_{\rm  m} = 0.2$, and the normalization of the matter power spectrum, $\sigma_8$, which essentially measures how clumpily matter is distributed, was determined to be $\sigma_8 = 0.838$, keeping all other cosmological parameters fixed \citep{ESJL}. Fig.~\ref{UnblindedBestFit} illustrates that there is no clear signal region from which the value of these parameters could have been read off. It simply depicts two correlation functions, $\xi_+$ and $\xi_-$, as a function of angular separation $\vartheta$.

This illustrates aptly how indirect astronomical constraints of physics often are. In fact,  they are often true `inference problems': Given the data, one wishes to infer by definition \emph{unobservable} quantities, namely the physical parameters. Only upon adoption of a likelihood can information on the parameters be distilled from the data.

In contrast to their inferences being extremely indirect, most astronomical raw data are easy to visualize and memorize. Salting, i.e. adding artificial signals to the data, is hence correctly disregarded in many astronomical disciplines due to artificial additions or omissions being easily spotted.

Astronomical and particle physics measurements are therefore close to opposites of each other, and a astronomical blinding technique needs to reflect this. We therefore describe an up to three-stage blinding strategy which may blind the likelihood, the theoretical predictions, and one of the last stages in the long and weary transition from raw data to science-ready data.

An example of this strategy is seen in Fig.~\ref{HolyBlindBestFit}, which uses the same data as Fig.~\ref{UnblindedBestFit}, only that the covariance matrix contained biases to shift the posterior. Overplotted is the blinded best-fit of $\Omega_{\rm m} = 0.274, \sigma_8 = 0.754 $, which differs by 1.5 posterior standard deviations from the actual best fit. This corresponds to a shift towards higher $S_8 = \sigma_8\sqrt{\Omega_{\mathrm m}/0.3}$, in a direction perpendicular to the degeneracy between $\Omega_{\mathrm m}$ and $\sigma_8$. Telling Fig.~\ref{HolyBlindBestFit} and Fig.~\ref{UnblindedBestFit} apart is extremely difficult, which underlines the power of the blinding algorithm now to be presented.

\subsection{Three-stage blinding setup}
\label{sec::three}
Although the below arguments are easily extended, we now specialize to Gaussianly distributed data. The likelihood is then Gaussian if the covariance matrix is known. If the covariance matrix is estimated from simulations, then the likelihood is the t-distribution of \citet{SH15,SH17} instead. These two cases are the most encountered likelihoods.

Parameter inference then requires three ingredients: the correct science-ready data $\fatx$, the correct covariance matrix $\boSig$ (either estimated or analytically computed), and the capability to compute the theoretical mean $\bomu(\both)$ at full accuracy, where $\both$ are the parameters of interest. It is thus natural to introduce three blinders, and we denote (potentially) blinded quantities with a breve, $\breve{\fatx},\breve{\boSig},\breve{\bomu}(\both)$.

We assume that to the general researcher, all three quantities $\breve{\fatx },\breve{\boSig},\breve{\bomu}(\both)$ appear blinded. To each of the three blinders, two out of three quantities appear blinded, and the third is the one whose blinding is their task.

The data-blinder requires access to the process of distilling science-ready data $\fatx$ from the raw data. The theory-blinder requires access to the software computing $\bomu(\both)$. Essentially any analysis at some point uses a look-up table, an emulator, or fixed nuisance parameters, all of which can be biased. The theory-blinder may  decide to restrict the other blinders' capabilities to compute $\breve{\bomu}(\both)$ -- this may be advantageous if blinding is partially assigned to external researchers.

Finally, we assume the likelihood-blinder has exclusive access to the code which computes the covariance matrix. This code is of no direct use as long as there is no data vector yet.
The likelihood-blinder is then given $\breve{\fatx}$, and computes its true covariance matrix. Additionally, the likelihood-blinder is given access to the potentially restricted  computational facilities to generate $\breve{\bomu}(\both)$. This blinder then uses the below algorithm to generate a biased covariance matrix, which is used in the evaluation of the blinded posterior. 

The blinding of $\breve{\fatx}$ and $\breve{\bomu}(\both)$ will be highly field specific. Many disciplines may even chose not to blind $\fatx$. Therefore, this paper now focuses on an algorithm to blind $\boSig$. The (potentially) blinded data and theory will then be re-encountered when jointly deblinding in Sect.~\ref{sec::deblinding}.

\section{Blinding preparations}
\label{sec::cov}
\subsection{Why blinded covariance matrices shift posteriors}
Having specialized to a Gaussian or $t$-distribution likelihood, blinding the likelihood is akin to blinding a covariance matrix. We thus have to explain why biases in a covariance matrix shift posteriors. This was partially discussed in \citet{ESJL}, which we here extend by showing how best-fitting parameters of a Gaussian likelihood directly depend on the covariance matrix.

We adopt a $d$-dimensional data vector $\fatx$, and indicate expectation values by angular brackets $\langle \cdot \rangle$. The mean is $\langle \fatx \rangle = \bomu = (\mu_1,...,\mu_d)$, which is a function of a $p$-dimensional parameter vector $\both = (\theta_1,...,\theta_p)$. The parameters are of physical significance and shall be inferred.

In general, $\bomu(\both)$ will depend non-linearly on its parameters, but to illustrate the biasing effect of covariance matrices we Taylor expand around an initial parameter point $\both_{\rm I}$. To this end, we introduce a non-square matrix $\sfX$ whose $j$th column shall be the derivative of the mean with respect to the $j$th parameter,
\begin{equation}
    \sfX = \left( \frac{\partial \bomu}{\partial \theta_1},...,\frac{\partial \bomu}{\partial \theta_r}  \right).
\end{equation}
The matrix $\sfX$ is thus $p\times d$ dimensional and its $ij$th element is given by
\begin{equation}
    X_{ij} = \frac{\partial \mu_i}{\partial \theta_j}.
\end{equation}
This matrix can be computed with any conventional Fisher matrix forecasting code \citep{Tegmark:1996bz,Class1,DALI,DALII}.
The linearized mean is then
\begin{equation}
\bomu_{\rm lin}(\both) = \bomu(\both_{\rm I}) + \sfX(\both - \both_{\rm I}),
\label{Taylor}
\end{equation}
which replaces the non-linear dependence on $\both$ by a linear dependence. In order for Eq.~(\ref{Taylor}) to hold, the derivatives in $\sfX$ must be evaluated at $\both_{\rm I}$. The aim is to determine the best-fitting parameters $\hat{\both}$, which maximize the log-likelihood. We adopt a Gaussian posterior with parameter-independent covariance matrix $\boSig$
\begin{equation}
\calP(\both|\fatx, \boSig) \propto \exp\left( -\half    \left[\fatx - \bomu(\both)\right]^\top \boSig^{-1} \left[\fatx - \bomu(\both)\right] \right),
\label{Gauss_x}
\end{equation}
where the superscript $\top$ denotes transposition. Replacing $\bomu(\both)$ with $\bomu_{\rm lin}(\both)$ it follows that the linearized best-fitting parameters must solve
\begin{equation}
    \frac{\partial}{\partial \both} \left[\fatx - \bomu_{\rm lin}(\both) \right]^\top \boSig^{-1} \left[\fatx - \bomu_{\rm lin}(\both)\right] = 0.
    \label{ToSolve}
\end{equation}
Using the relation
\begin{equation}
    \frac{\partial }{\partial \boldsymbol{s}}(\boldsymbol{r} - \sfA\boldsymbol{s})^\top \sf{\Omega}(\boldsymbol{r} - \sfA\boldsymbol{s}) = -2 \sfA^\top\Omega(\boldsymbol{r} - \sfA\boldsymbol{s}),
\end{equation}
for vectors $\boldsymbol{s},\boldsymbol{r}$ and matrices $\sfA, \sf{\Omega}$ of matching dimensions, the solution to Eq.~(\ref{ToSolve}) yields the linearized best-fitting parameters
\begin{equation}
    \hat{\both}_{\rm lin} = (\sfX^\top \boSig^{-1}\sfX)^{-1} \sfX^\top \boSig^{-1} [\fatx - \bomu(\both_{\rm I}) + \sfX\both_{\rm I}].
    \label{BF}
\end{equation}
This illustrates that the position of the best-fit depends on the covariance matrix in a dual manner: the term
\begin{equation}
    \boSig^{-1} [\fatx - \bomu(\both_{\rm I}) + \sfX\both_{\rm I}],
\end{equation}
inverse-variance weights the distance between the data and the mean, thereby preferring means which match the data in units of the covariance. The term
\begin{equation}
     (\sfX^\top \boSig^{-1}\sfX)^{-1},
     \label{compression}
\end{equation}
describes a compression, due to $\sfX$ being non-square. This term compresses the information from fitting in data space into the lower-dimensional parameter space.

As the best-fitting parameters depend in this dual manner on the covariance matrix, it directly follows that a bias in the covariance matrix will translate into a shift of the best-fitting parameters -- this opens the possibility to blind by changing the covariance matrix.

In fact, for the purpose of blinding, the compression term of Eq.~(\ref{compression}) has a further appeal: due to the compression, many different covariance matrices will lead to the same shift in parameters, due to this being a highly underdetermined system. This allows us to set side-constraints, for example that the biased covariance matrix not only induces a specified shift from best-fitting parameters $\hat{\both}$ to blinded best-fitting parameters $\breve{\both}$, but at the same time also maintains e.g.~all its original variances, and (as a further example) its determinant, or the sign of all its correlation coefficients.

The upcoming sections therefore describe blinding by constructing one (or multiple) blinded covariance matrices. To linear order, the blinded best-fit will then lie at 
\begin{equation}
      \breve{\both}_{\rm lin} = (\sfX^\top \breve{\boSig}^{-1}\sfX)^{-1} \sfX^\top \breve{\boSig}^{-1} [\fatx - \bomu(\both_{\rm I}) + \sfX\both_{\rm I}],
\label{linbias}
\end{equation}
which will be extended to a fully non-linear inference with sampling in Sect.~\ref{sec::blindpost}.

\begin{table*}
\caption{$\Delta \chi^2$-values where the 90\% credibility contour lies above the minimum $\chi^2$, as a function of number of parameters. If the blinder chooses target parameters $\both_t$ which differ from the initial parameters $\both_o$  by more than $\Delta \chi^2$ shown here, then a posterior shift of more than three posterior standard deviations will ensue. This is not advised.}
\label{tab:example}
\begin{tabular}{lcccccccccccccccccc}
\hline
parameters   & 6  & 8  & 10 & 12 & 14 & 16 & 18 & 20 & 22 & 24 & 26 & 28 & 30 & 40 & 50 & 100\\
\hline
$\Delta \chi^2$ (90\%) & 10.6  & 13.3 & 15.9 & 18.5 & 21.0 & 23.5 & 25.9 & 28.4 &30.8 & 33.2 & 35.5 & 37.9 & 40.2 & 51.8 & 63.1 & 118.5 \\
\hline
\label{Tab1}
\end{tabular}
\end{table*}

\subsection{Stages of the covariance blinding algorithm}
\label{sec::blind}
The upcoming blinding algorithm passes through different stages. At first, the algorithm assists the blinder in determining sensible magnitudes for the parameter shift (Sect.~\ref{sec::target}). This is non-trivial, as the posterior width is not yet known. The algorithm then transforms the data from their physical units onto a representation which is natural for statistical manipulations (Sect.~\ref{sec::datatrafo}). Subsequently, Sect.~\ref{sec::blindalg} describes how to adapt the covariance matrix to induce the intended posterior shift. 

Theoretically, the algorithm could then stop. However, wilful deblinding might at that stage still be possible, and the entirety of Sect.~\ref{sec::disable} is thus devoted to making wilful deblinding impossible by using an encryption algorithm and allowing the specification of side constraints to be met. Sect.~\ref{sec::control} computes the shift of parameters to determine whether blinding succeeded.
Finally, the physical units are restored, and the output is a purposefully biased covariance matrix which leads to the requested bias in physical parameters, with the origin of the bias being untraceable due to the intermediate encryption and side constraints.

\subsection{Finding sensible target parameters for blinding}
\label{sec::target}
At the outset of the blinding strategy, the blinder has to pick an origin $\both_{\rm o} $ of the posterior shift, and target parameters $\both_{\rm t}$ which are preferred after shifting. Typically, shifting the yet unknown posterior by two to three standard deviations is the sought aim. Having never computed the full posterior, its standard deviations are however not known yet, so we effectively wish to shift by the multiple of a yet unknown quantity. We thus require an abstract prediction of the shift's magnitude.

To this aim, we exploit scaling relations: For a multivariate Gaussian posterior, the best-fit will occur at the minimal chisquared, $\chi^2_{\rm min}$. The associated 90-percent credibility contour can then be chosen to run along an isocontour which lies by $ \Delta \chi^2$ above this minimum. The more parameters are estimated, the larger $\Delta \chi^2$ has to be, since more parameters will widen up the joint posterior. The values of $\Delta \chi^2$ where the 90-percent credibility contour lies above the minimum are known (they are integrals of $\chi^2$-distributions), and we tabulate them as a function of free parameters in table~\ref{Tab1}.

The blinder thus has to know how many parameters are to be inferred, and has to pick initial parameters $\both_{\rm o}$ and target parameters $\both_t$ which differ approximately by $\Delta \chi^2$-values as given in table~\ref{Tab1}. 

If the blinder has no physical intuition for which parameter values $\both_{\rm o}$ are likely to be preferred by the data, then the linearized best-fitting estimator of Eq.~(\ref{BF}) can be evaluated, which will yield parameter values $\hat{\both}_{\rm lin}$ close to the full non-linear best fit $\hat{\both}$.

We found that the values $\both_{\rm o}$ do not need to be picked with too much care, as long as they are within about two posterior standard deviation from the true best-fit $\hat{\both}$. We provide a public code\footnote{https://github.com/elenasellentin/StellarBlind} for the algorithm, which assists the blinder via  Eq.~(\ref{BF}) and table~\ref{Tab1} in finding sensible parameters $\both_{\rm o}$ for the shift origin, and for the shift destination $\both_t$.

\begin{figure*}
\includegraphics[width=\textwidth]{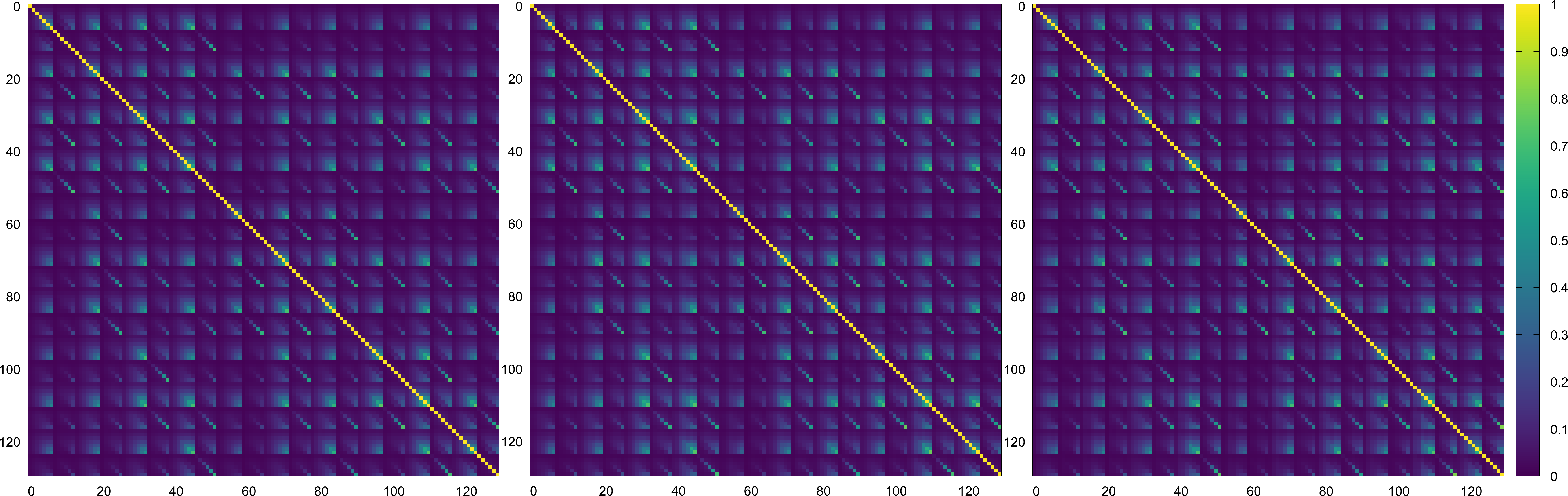} 
\caption{Three correlation matrices, of which one is the original correlation matrix of \citet{KiDS450}, and two were constructed such that the posterior peaks at pre-defined parameter values. These matrices result in the three posteriors of Fig.~\ref{KiDS-450Posterior}. The extreme difficulty of identifying the correct correlation matrix illustrates the power of the here presented blinding algorithm.}
\label{Spot}
\end{figure*}

\subsection{Data transformation}
\label{sec::datatrafo}
The general astronomical data set will come in `natural' units, which might depend on estimator choice or the adopted (physical) units of measurements, such as parsec or mega-parsec. To gain generality, we thus standardize the data set.

Let $\Sigma_{ii}$ be the $i$th diagonal element of the covariance matrix, then the conditional standard deviations $\sigma_i$ are
\begin{equation}
\sigma_i = \sqrt{\Sigma_{ii}},
\label{sigsis}
\end{equation}
The standard deviations have the same dimensions as the data points themselves. We therefore introduce the standardized variables
\begin{equation}
u_i = \frac{x_i}{\sigma_i}, \ \ \nu_i = \frac{\mu_i}{\sigma_i},
\label{standard}
\end{equation}
such that $\boldsymbol{u}$ is the data vector expressed as multiples of its former standard deviation, and $\boldsymbol{\nu}$ is the standardized theoretical mean, which still depends on parameters $\both$. Dividing out the standard deviations at this point allows us to restore them post-blinding. This is required for situations where variances can be easily remembered, and must thus remain unchanged during blinding.

The covariance matrix of the standardized data is then the correlation matrix
\begin{equation}
\left\langle (\bou - \langle\boldsymbol{u}\rangle)(\bou - \langle\boldsymbol{u}\rangle)^\top\right\rangle = \sfC.
\end{equation}
The posterior of the parameters given the standardized data is then a Gaussian, with the inverse correlation matrix as precision matrix
\begin{equation}
\calP(\both | \bou) \propto \exp\left( -\half    \left[\bou - \boldsymbol{\nu}(\both)\right]^\top\invsfC \left[\bou - \boldsymbol{\nu}(\both)\right] \right).
\label{Gauss_u}
\end{equation}
Eq.~(\ref{Gauss_u}) and Eq.~(\ref{Gauss_x}) are exactly the same posterior, only once expressed in units natural for statistics (Eq.~\ref{Gauss_u}), and once in units natural for the astronomer (Eq.~\ref{Gauss_x}).

Finally, we Cholesky decompose the inverse correlation matrix
\begin{equation}
    \invsfC = \sfL \sfL^\top,
\end{equation}
where $\sfL$ is a lower triangular matrix, and its transpose is upper-triangular.

Technically, $\sfL^\top \bou$ is a whitening transform, with the astronomical implication being that $\sfL$ causes the often rich structure of astronomical data. Factorizing it out at this point enables us to multiply it back in later, whereupon deceivingly naturally-looking covariance matrices are restored.

\section{Main blinding algorithm}
\label{sec::blindalg}
At this stage, the data have been transformed into a easily manageable representation, and the blinder has decided which shift of the posterior shall be induced. We therefore now lay out the mathematics of how to change a covariance matrix such that it shifts the posterior in a wanted direction. 

To clarify the aim, we depict in Fig.~\ref{Spot} three correlation matrices as produced with the upcoming algorithm. One of these is the original correlation matrix of \citet{KiDS450}, and two are matrices biased by our algorithm. When used in a Gaussian likelihood, the three matrices lead to the three posteriors of Fig.~\ref{KiDS-450Posterior}. Obviously, maximally one of the three posteriors can be correct. To underline the point of how easily such shifts are hidden in a covariance matrix, we refrain from revealing which of the three matrices in Fig.~\ref{Spot} is the correct correlation matrix.\footnote{In the spirit of reproducible research, all plots in this paper can be reproduced with our public code and the equally public likelihood of \citet{KiDS450}.} The algorithm for blinding the covariance matrix is as follows.

\subsection{Biasing the covariance matrix}
\label{sec::inject}
The algorithm begins by translating the posterior, and side constraints are enforced later.

To achieve a translation, we have to adapt the $\chi^2$-surface. Expressed by the Cholesky decomposition, the chisquare surface is
\begin{equation}
 \chi^2(\both,\bou, \sfL  ) = \left[\bou - \boldsymbol{\nu}(\both)\right]^\top \sfL \sfL^\top \left[\bou - \boldsymbol{\nu}(\both)\right].
\end{equation}
We now introduce the biased inverse correlation matrix $\breve{\sfC}^{-1}$, for which we chose the ansatz 
\begin{equation}
\breve{\sfC}^{-1} = \sfL \sfB\sfB^\top \sfL^\top,
\label{ansatz}
\end{equation}
where $\sfB = \mathrm{diag}(b_{11},b_{22},...,b_{dd})$ is a diagonal matrix which causes the translation. The matrix $\sfB$ can be interpreted as artificial signals, which are hidden in the correlation matrix. Eq.~(\ref{ansatz}) also implies that this blinding technique requires a dense correlation matrix: If the original correlation matrix were diagonal, then $\sfL$ were diagonal as well, and the blinding would then easily be discovered.

To compute $\sfB$ such that the posterior shifts from $\both_{\rm o}$ to $\both_{\rm t}$, we follow the arguments of \citet{ESJL} and demand $\chi^2(\both_{\rm o},\bou, \sfC) = \chi^2(\both_{\rm t}, \bou, \breve{\sfC})$ which requires
\begin{equation}
\begin{aligned}
 & \left[\bou - \boldsymbol{\nu}(\both_{\rm o})\right]^\top \sfL \sfL^\top \left[\bou - \boldsymbol{\nu}(\both_{\rm o})\right] \\
  = & \left[\bou - \boldsymbol{\nu}(\both_{\rm t})\right]^\top \sfL \sfB \sfB^\top \sfL^\top \left[\bou - \boldsymbol{\nu}(\both_{\rm t})\right].
 \end{aligned}
 \label{constraint}
\end{equation}
This constraint expresses that the $\chi^2$-value prior to blinding at the origin shall equal the $\chi^2$-value post-blinding at the parameters targeted by blinding. As $\sfB$ is diagonal, it can now easily be computed. We introduce the vectors
\begin{equation}
\begin{aligned}
    \sfL^\top [\bou - \boldsymbol{\nu}(\both_{\rm o})] & = \boldsymbol{e},\\
    \sfL^\top [\bou - \boldsymbol{\nu}(\both_{\rm t})] & = \breve{\boldsymbol{e}},
\end{aligned}
\label{evectors}
\end{equation}
and the diagonal elements of $\sfB$ are then
\begin{equation}
    b_{ii} = \frac{e_i}{\breve{e}_i}. 
\end{equation}
Theoretically, the blinding algorithm could stop here: one could now directly compute $\breve{\sfC} = (\sfL \sfB \sfB^\top \sfL^\top)^{-1}$, which is the blinded correlation matrix. The variances which were divided out in Eq.~(\ref{standard}) would need to be multiplied back in, and the result would be a blinded covariance matrix which shifts the posterior. 

However, at this point human deblinding might still be possible as the above is a deterministic calculation. There might exist situations where enough intuition about the true covariance matrix can be gained in order to reverse-engineer which blinding parameters $\both_{\rm t}$ the blinder chose. The blinding could then be undone.

Additionally, because the determinant $|\sfB|$ was not enforced to be unity, we will have changed the determinant of the covariance matrix. This will change the size of the posterior.

We therefore regard the ansatz Eq.~(\ref{ansatz}) with a diagonal $\sfB$ only as a convenient starting point for the algorithm, and we will now exploit the fact that the constraint of Eq.~(\ref{constraint}) is strongly underdetermined: it sets a single constraint to solve for the elements of a (in general dense) $d\times d$ matrix. There thus exist infinitely many matrices $\sfB$ to induce the wanted posterior translation, and in the following we use this freedom to adapt the blinded correlation matrix.

\section{Disabling accidental deblinding}
\label{sec::disable}
The former section Sect.~\ref{sec::blindalg} biased a covariance matrix such that the posterior prefers a chosen set of parameters $\both_{\rm t}$. The entirety of this section is devoted to making the blinding untraceable. As this requires us to employ random manipulations, this section ends by controlling whether the requested shift is still achieved, despite the manipulations.

\subsection{Disabling recovery of the blinding parameters}
We improve the quality of the algorithm by making it impossible to reconstruct the parameters chosen by the blinder. This requires us to change the values $b_{ii}$, subject to still inducing the wanted shift. To this aim, we note that the magnitude of the $b_{ii}$ will in general be too large, and it is preferable if the $b_{ii}$ are as close to unity as only possible during blinding. 

To compute how close to unity we can push the $b_{ii}$, we compute the average $\Delta \chi^2$ that the blinding must induce. Using the vectors from Eq.~(\ref{evectors}), the $\chi^2$ at the target parameters is
\begin{equation}
 \chi^2(\both_{\rm t},\bou,\sfC) = \breve{\boldsymbol{e}}^\top  \breve{\boldsymbol{e}},
\end{equation}
before blinding. This will be larger than the $\chi^2$ after blinding, which is
\begin{equation}
 {\chi}^2(\both_{\rm t},\bou,\breve{\sfC}) = \breve{\boldsymbol{e}}^\top \sfB \sfB^\top  \breve{\boldsymbol{e}}.
\end{equation}
The $\Delta \chi^2$ bridged during blinding is thus
\begin{equation}
\begin{aligned}
 \Delta \chi^2 & = \chi^2(\both_{\rm t},\bou,\sfC) - {\chi}^2(\both_{\rm t},\bou,\breve{\sfC}) \\
 & = \sum_{i = 1}^d \breve{e}_i^2 [1-b_{ii}^2]. 
 \end{aligned}
 \label{c}
\end{equation}
It is senseless to fit perfectly to one realization $\breve{\boldsymbol{e}}$ of the data, and we thus average over multiple realizations. On average, we will have $\langle \breve{e}_{ii}^2\rangle \approx 1$ because $\breve{\boldsymbol{e}}$ is approximately a white vector, where the approximation is that blinding conditions on an incorrect mean, see Eq.~(\ref{evectors}).

\begin{figure}
\includegraphics[width=0.45\textwidth]{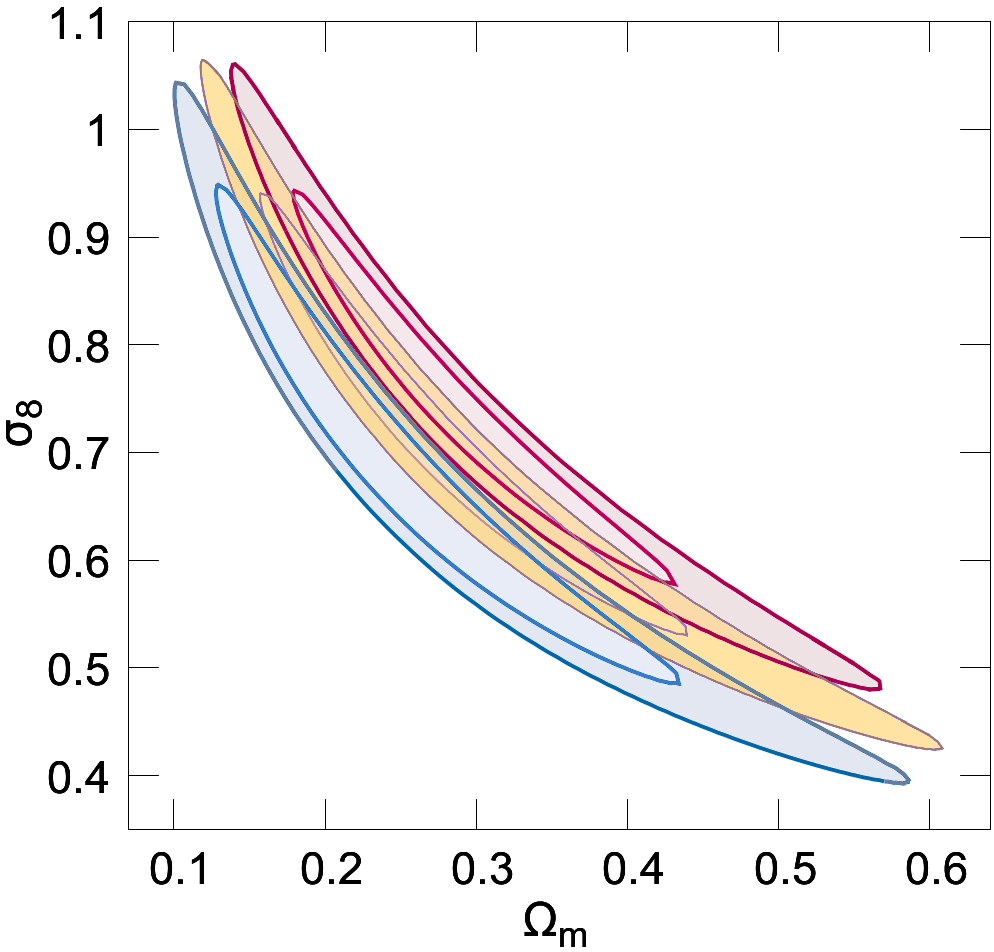} 
\caption{Blinded analysis of the KiDS-450 data vector, once with the true covariance matrix (orange), and once with two blinded covariance matrices, whose corresponding correlation matrices are depicted in Fig.~\ref{Spot}.}
\label{KiDS-450Posterior}
\end{figure}

As a consequence of $\breve{\boldsymbol{e}}$ being approximately white it follows that all $b_{ii}$ are approximately the same. Then, inserting $\langle \breve{e}_{i}^2 \rangle = 1$ into Eq.~(\ref{c}), we have
\begin{equation}
 \langle \Delta \chi^2\rangle \approx d(1-b_{ii}^2).
\end{equation}
Solving for the elements $b_{ii}$, we have
\begin{equation}
 |b_{ii}| = \sqrt{ 1- \frac{\langle \Delta \chi^2 \rangle  }{d} }.
\end{equation}
This result has a highly intuitive interpretation: if blinding bridges a larger gap $\Delta \chi^2$, then the magnitude of the $b_{ii}$ increases, implying larger biases are needed. On the other hand, if the dimension $d$ of the data increases, then smaller elements $b_{ii}$ suffice to still shift the posterior. For $d\to\infty$, we have $b_{ii} \to 1$, meaning even the tiniest changes in the covariance matrix will suffice to induce major shifts of the posterior. Accordingly, it is easier to blind large data sets.

For our blinding algorithm, we hence set
\begin{equation}
\forall \ i : \  \sqrt{ 1 - K } <  b_{ii} < \sqrt{ 1 + K }.
\end{equation}
These thresholds will deteriorate the goodness of fit, which we will compensate for later. The scalar $K$ is
\begin{equation}
 K = W \frac{\chi^2(\both_{\rm t},\bou,\sfC) - {\chi}^2(\both_{\rm t},\bou,\breve{\sfC})}{d},
 \label{room}
\end{equation}
where $W$ has to be positive and acts as a tolerance: if $W<1$, then blinding will not succeed, as the $b_{ii}$ can vary insufficiently to induce the requested change in chisquared. If $W>1$, then the $b_{ii}$ can induce even larger changes in $\Delta \chi^2$ than is needed to shift the posterior. The latter allows further constraints to be enforced.

We now additionally demand the determinant $|\sfB|$ be unity. We thus
rescale all elements $b_{ii} \rightarrow b_{ii} |\sfB|^{-1/d}$, where $|\sfB|$ is the determinant of $\sfB$. After rescaling, the determinant of $\sfB$ is unity, which will not alter the determinant of the blinded correlation matrix. This will in turn leave the size of the posterior unchanged.

The elements $b_{ii}$ have now been changed twice: first their magnitude has been subjected to upper and lower bounds, and then all elements were jointly changed multiplicatively. Both will induce degeneracies in attempts of recovering the original $b_{ii}$ and the blinder's $\both_{\rm t}$, thereby making successful reverse engineering very unlikely. To disguise the biases even further, we shall now hide their presence through a series of random changes and matrix inversions.

\subsection{Encrypting the correlation matrix}
\label{sec::encrypt}
At this stage, we multiplied a biasing matrix to the Cholesky decomposition of the inverse correlation matrix
\begin{equation}
    \breve{\sfL}^\top = \sfB^\top \sfL^\top,
\end{equation}
where $\sfB$ has already been pre-optimized. The biased inverse correlation matrix would then be $\brinvsfC = \brsfL\brsfL^\top$, but in Eq.~(\ref{room}), we left margin to optimize further ($W>1$), which we now use. 

We now transit to random manipulations, such that as long as the random seeds of these random manipulations are unknown, they cannot be reversed. The below algorithm can be seen as an encryption, where the random seeds are the decryption keys.

We introduce the `symmetrized mean absolute percentage error' (SMAPE) between two values $v,v'$, which is given by
\begin{equation}
    \mathrm{SMAPE}(v,v') = \frac{|v-v'|}{|v|+|v'|}. 
    \label{eq::smape}
\end{equation}
This symmetrized percentual error yields zero if $v'=v$. Its upper bound of unity is reached in the limits $v\gg v'$ and $v' \gg v $. For $v' = 2v$ it yields $1/3$, so it scales differently from the usual percentual error. The latter is a desired benefit: during blinding, elements of either the blinded or unblinded matrices can become zero, which the symmetrized error handles well, whereas the usual percentual error would return misleading zeros or infinite values.

To disguise the presence of the biases, we now shrink $\breve{\sfL}$ element wise randomly towards the unbiased ${\sfL}$. The diagonal elements are left unchanged, in order to not change the determinant. Thus, 
\begin{equation}
\begin{aligned}
 \forall \ i<j: \ & {\rm if\ }  \mathrm{SMAPE}( \breve{\sfL}_{ij}, \sfL_{ij} ) > S_{\rm inv},\\
                  & \mathrm{then} \ \breve{\sfL}_{ij} \sim \ \mathrm{Uniform}(\breve{\sfL}_{ij}, {\sfL}_{ij}).
 \end{aligned}
\end{equation}
Here, $S_{\rm inv} \in [0,1]$ is a user defined threshold for the SMAPE. If the elementwise SMAPE is exceeded, then the element of $\breve{\sfL}_{ij}$ is reset to resemble the unbiased element ${\sfL}_{ij}$ more closely, by randomly drawing from a uniform distribution with upper and lower bound given by the two matrix elements. For this random draw, a seed has to be specified.

Given the elementwise edited matrix $\breve{\sfL}$, we compute $\brinvsfC = \brsfL\brsfL^\top$, and invert it to yield $\breve{\sfC}$. This inversion redistributes the random changes in a manner impossible to predict by humans. To edit even further, we now Cholesky decompose the blinded and unblinded correlation matrices
\begin{equation}
\begin{aligned}
\sfC & = \sfR \sfR^\top,\\
 \breve{\sfC} & = \breve{\sfR} \breve{\sfR}^\top,
 \end{aligned}
\end{equation}
and we repeat the shrinking towards the original Cholesky decomposition
\begin{equation}
\begin{aligned}
 \forall \ i<j: \ & {\rm if\ }  \mathrm{SMAPE}( \breve{\sfR}_{ij}, \sfR_{ij} ) > S_{\rm corr},\\
                  & \mathrm{then} \ \breve{\sfR}_{ij} \sim \ \mathrm{Uniform}(\breve{\sfR}_{ij}, {\sfR}_{ij}).
 \end{aligned}
\end{equation}
Here, $S_{\rm corr}$ is a threshold for the SMAPE which specifies the maximal changes the blinder tolerates in elements of the Cholesky decomposition of the correlation matrix. 

The random resetting here conducted partially erases the desired biases, which is why it was important to leave room in Eq.~(\ref{room}). Should the partial erasing cause the blinding control of Sect.~\ref{sec::control} to fail, then $W$ in Eq.~(\ref{room}),  $S_{\rm inv}$ and $S_{\rm corr}$ have to be adjusted.  Crucially though, it is by this stage impossible to reconstruct how the biases entered the correlation matrix -- two random editing processes with an intermediate inversion lie in the way.

\subsection{Optimization of further side-constraints}
\label{sec::sideconstraints}
The editing process of the biased correlation matrix has meanwhile progressed so far that the biased correlation matrix will strongly resemble the unblinded matrix. This makes it easy to optimize for final constraints which may be necessitated by the specifics of the research field -- this step is highly important, as it is the only one to enforce that all physical constraints are met. Should logical inconsistencies remain and be discovered during the blinded analysis, then fractions of the blinding might become reversible.

For example, one might wish that the variances do not change during blinding, or that the correlation coefficients do not change sign, if there is a physical reason for positive or negative correlation between data points. Other fields might require the eigenvalues to be unchanged. If independent experiments are combined, then a block-diagonal structure of the covariance matrix ensues, which also ought to be preserved during blinding. Almost certainly one might wish that the formerly preferred parameter point $\both_{\rm o}$ is now disfavoured with at least a certain $\Delta \chi^2$ with respect to the target parameters $\both_{\rm t}$. 

If any such constraints has to be perfectly fulfilled, then the blinder should enforce it directly, e.g. by resetting the variances, and transit to controlling the success of the blinding algorithm in Sect.~\ref{sec::control}. But in general, enforcing additional constraints without major care may corrupt the correlation matrix. For example, resetting variances may lead to a non-positive definite matrix. Instead of implementing any constraints by brute-force, we rather advocate the following stochastic gradient descent algorithm, which operates on Cholesky decompositions instead.

We define a loss function $F$ which is the sum over all constraints, such as
\begin{equation}
\begin{aligned}
    F & = \mathrm{SMAPE}(\breve{\sfL},\sfL) + \mathrm{SMAPE}(\breve{\sfR},\sfR) \\
    & + \left[\chi^2(\both_{\rm o},\bou,\breve{\sfC}) - \chi_{\rm req}^2(\both_{\rm o})  \right]^2 \\
    & + \left[\chi^2(\both_{\rm t},\bou,\breve{\sfC}) - \chi_{\rm req}^2(\both_{\rm t})  \right]^2 \\
    & + \sum_i (\breve{C}_{ii} - C_{ii})^2,
    \end{aligned}
\end{equation}
where we define the SMAPE of a matrix to be taken elementwise, $\chi^2$-values without subscript are those achieved when using the current iteration's matrix, and $\chi_{\rm req}^2$ with subscript are numbers which are the blinder's requested values at these parameter points. Omitting or adding further constraints to the loss function is possible until an over-constrained system is reached.

The loss function $F$ now has to be minimized. We found a particularly efficient minimization alternately changes random elements of $\breve{L}$ and $\breve{R}$ on the few percent level. If the loss $F$ decreased, the random change is accepted and the iteration proceeds to changing new matrix elements. 

If the loss did not decrease, the random change is discarded without updating the current matrices $\breve{\sfL}$ and $\breve{\sfR}$, and a new iteration is begun. 

The minimization of $F$ can be stopped when the blinder's targets are reached, or if $F$ begins to asymptote to the minimal loss achievable under the set constraints. In practise, the loss function must include constraints on disfavouring the old parameters $\both_{\rm o}$ with respect to the target parameters $\both_{\rm t}$, otherwise minimizing the loss function will reproduce the unblinded correlation matrix.

\subsection{Setting the target chisquare}
Former experience with blinding of \citet{KiDS450} revealed that out of a set of blinded posteriors, many researchers suggested the one with the smallest minimum-$\chi^2$ represents the true posterior. This is an incorrect assumption, and indeed turned out to be wrong for \citet{KiDS450}. To counter this pre-conception, we suggest the blinder also enforce a $\chi^2$-value of their choice.

This is easily achieved, e.g.~through the loss function of Sect.~\ref{sec::sideconstraints}. Another possibility, which will change the determinant, is to reset the eigenvalues of the biased correlation matrix. In this case, the biased correlation matrix is to be spectrally decomposed $\breve{\sfC} = \sfQ \sfLam \sfQ^{-1}$, where $\sfQ$ is an orthogonal $d\times d$ matrix satisfying $\sfQ^\top = \sfQ^{-1}$. The matrix $\sfLam$ is the diagonal matrix of eigenvalues. The chisquared value at the target parameter point is thus
\begin{equation}
\begin{aligned}
    \chi^2(\both_{\rm t}) & = \left[\bou - \boldsymbol{\nu}(\both_{\rm t})\right]^\top \left[ \sfQ \sfLam \sfQ^{-1} \right]^{-1} \left[\bou - \boldsymbol{\nu}(\both_{\rm t})\right]\\
    & = \sum_{i} \frac{1}{\lambda_i} \left( \boldsymbol{q}_i^\top [\bou-\boldsymbol{\nu}(\both_{\rm t})] \right)^2.
    \end{aligned}
    \label{evals}
\end{equation}
Here, $\lambda_i$ is the $i$th eigenvalue and the vectors $\boldsymbol{q}_i$ are the $i$th row of the matrix $\sfQ^{-1}$.

From Eq.~(\ref{evals}) we thus see that the eigenvalues weight the contribution of each summand to the total $\chi^2$. The blinder may thus reset either a single, or multiple eigenvalues to enforce the $\chi^2$ of their choice at target parameters $\both_{\rm t}$.

\begin{figure}
\includegraphics[width=0.47\textwidth]{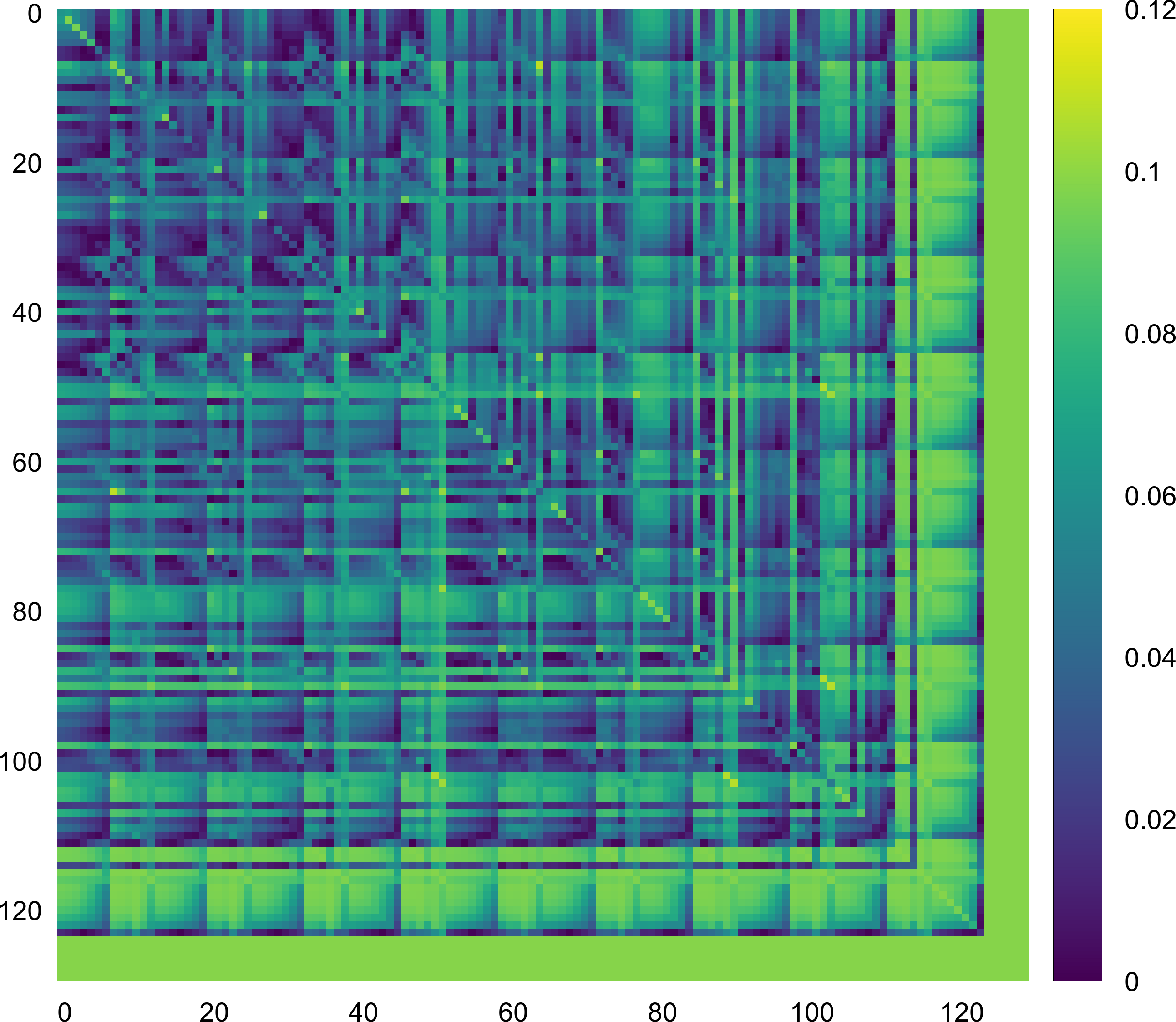} 
\caption{Plot of the relative changes each matrix element of the covariance matrix underwent whilst blinding the KiDS-450 analysis. The colour bar measures the `symmetrized mean absolute percentage error' (SMAPE), defined in Eq.~(\ref{eq::smape}). The here plotted relative difference matrix causes the upwards shift from orange to red posterior in  Fig.~\ref{KiDS-450Posterior}. The green bars of SMAPE values around 0.1 for matrix rows 120-130 affect the highest redshift bin of KiDS-450, and there preferentially $\xi_-$. This implies that the upwards shift of the KiDS-450 posterior can be caused by mischaracterizing the correlation of the data from the highest redshift bin with all lower redshift bins. Appendix \ref{Kids_specific} provides further cosmic shear specific context for this figure.}
\label{SmapeCovs}
\end{figure}

\subsection{Finalization of the algorithm and output}
At this stage, the blinding algorithm has nearly completed, with the current output being the prototype $\breve{\sfC}$ of the biased correlation matrix. Importantly though, the algorithm does not enforce the diagonal elements of the correlation matrix to be unity during blinding. This is fully acceptable and simply corresponds to a rescaling of the variances. The final step is thus to transform back to the natural units of the astronomical data, which yields the final biased covariance matrix and the final biased correlation matrix.

We therefore multiply back in the variances that were factored out in Eq.~(\ref{sigsis})
\begin{equation}
    \breve{\Sigma}_{ij} = \breve{C}_{ij}\sigma_i \sigma_j.
\end{equation}
This creates the final blinded covariance matrix $\breve{\boSig}$. Thus, if any of the $\breve{C}_{ii} \neq 1$ at this stage, then this simply rescales the variance $\breve{\Sigma}_{ii}$. The \emph{final} blinded correlation matrix will then nonetheless have unit diagonal elements. It results from the computation
\begin{equation}
    \breve{C}_{ij} = \frac{\breve{\Sigma}_{ij}}{\sqrt{\breve{\Sigma}_{ii} } \sqrt{\breve{\Sigma}_{jj} }  }.
\end{equation}
At this stage, the blinding algorithm is completed, with $\breve{\boSig}$ and $\breve{\sfC}$ being the final blinded covariance and correlation matrix. 

An example of the relative differences between the unblinded and final blinded covariance matrix is seen in Fig.~\ref{SmapeCovs}, from which it can indeed be seen that the variances (diagonal elements) changed somewhat, but that the posterior shift is mostly induced by having changed off-diagonal covariance elements. This also explains why spotting the blinding from the joint plot of data and theory curves in Figs.~\ref{UnblindedBestFit} and \ref{HolyBlindBestFit} is essentially impossible: the covariances do not show up when overplotting data and theory predictions, and thus go unnoticed when attempting to fit by eye.

Examples of final biased correlation matrices are seen in Fig.~\ref{Spot}, where it is difficult to spot the unbiased correlation matrix amongst the two biased correlation matrices. The corresponding biased and unbiased covariance matrices are equally difficult to tell apart, but due to the disadvantageous scaling over many order of magnitudes (see y-axes of the data in Fig.~\ref{UnblindedBestFit}) this is difficult to visualize in a colour plot.

\subsection{Blinding control}
\label{sec::control}
Although highly reliable, the algorithm does contain free algorithmic parameters $W,S_{\rm inv}, S_{\rm corr}$, an adaptable loss function $F$, and random seeds. It may thus sometimes fail, either due to a user error or due to chance. Before using the blinded covariance matrix $\breve{\boSig}$ in a posterior, the blinder has to control the sanity of the matrices, and that
the intended posterior shift was achieved.

The biased covariance matrix will be mathematically sound, if all variances are positive, and if the final biased correlation matrix is positive definite, and its elements take values on the interval $[-1,1]$. As the algorithm used Cholesky decompositions and spectral decompositions, positive definiteness should be guaranteed.

Whether the blinded covariance matrix will shift the posterior as intended can be evaluated in multiple ways. To linear order, it can be checked whether the best-fitting estimator Eq.~(\ref{linbias}) indeed yields parameters close to the targeted $\both_{\rm t}$. To non-linear order, it should be ensured that $\chi^2$ at the shift's origin has increased during blinding, i.e.
\begin{equation}
   \chi_{\rm final}^2(\both_{\rm o},\bou,\breve{\sfC}) > \chi^2(\both_{\rm o},\bou,\sfC),
\end{equation}
which expresses that the blinded covariance matrix disfavours the former parameters. Simultaneously, $\chi^2$ at the target parameters should have decreased due to blinding
\begin{equation}
   \chi_{\rm final}^2(\both_{\rm t},\bou,\breve{\sfC}) < \chi^2(\both_{\rm t},\bou,\sfC),
\end{equation}
expressing that the formerly disfavoured parameters are now a better fit than before.

To check whether the posterior shifted by the intended number of standard deviations, the final check is to compute the final $\Delta \chi^2$ between formerly preferred and target parameters
\begin{equation}
   \Delta \chi_{\rm final}^2 = \chi_{\rm final}^2(\both_{\rm o},\bou,\breve{\sfC}) - \chi_{\rm final}^2(\both_{\rm t},\bou,\breve{\sfC}).
\end{equation}
This $\Delta \chi_{\rm final}^2$ has to be compared against Tab.~\ref{Tab1}, in order to control whether its magnitude shifts by sufficiently many standard deviations for the total number of parameters to be fitted. Should this $\Delta \chi_{\rm final}^2$ be negative, then the old parameters were still preferred.

If any of these tests is failed, then the blinding algorithm has to be run with adapted algorithmic parameters. In the public code accompanying this paper, the code reports whether a test is failed, and recommends improved settings of the algorithmic parameters. Otherwise, if all tests are passed, then the yielded biased covariance matrix $\breve{\boSig}$ is qualified for use in a blinded posterior evaluation.

\section{Computing the blinded posterior}
\label{sec::blindpost}
The above algorithm allows the construction of biased covariance matrices $\breve{\boSig}$ which shift the posterior in a requested direction. As the algorithm has free parameters, including $S_{\rm inv}, S_{\rm corr}, \both_{\rm t}$ and also random seeds, many such covariance matrices can be computed. If multiple covariance matrices are computed, care should be taken that the entire set of covariance matrices does not allow joint deblinding, for example by averaging.

Any such biased covariance matrix would then be used to compute the blinded Gaussian posterior
\begin{equation}
    \breve{\calP}(\both|\breve{\fatx},\breve{\boSig}) \propto \exp\left( -\half    \left[\breve{\fatx} - \bomu(\both)\right]^\top \breve{\boSig}^{-1} \left[\breve{\fatx} - \bomu(\both)\right] \right),
\end{equation}
or in the blinded $t$-distribution of \citet{SH15}.
This posterior's peak position will be jointly influenced by the data-blinder's target parameters, and the likelihood-blinder's target parameters. An example of such shifted posteriors is seen in Fig.~\ref{KiDS-450Posterior}, where the original data $\fatx$ of KiDS-450 were used.

\section{Deblinding}
\label{sec::deblinding}
The essential step at the end of any blinded analysis is of course to deblind.\footnote{We distinguish between unblinded and deblinded. An unblinded quantity never was blinded, and a deblinded quantity was intermittendly blinded but the blinding is then undone.} In our setup, up to three quantities were blinded, the data, the theory computations, and the covariance matrix. The theory blinder is strongly recommended to deblind directly after having received the blinded data and the blinded covariance. This means the theory blinder is the only one who is recommended to deblind \emph{prior} to computing the blinded posterior. 

This recommendation has a utilitarian aim: the numerical costs of posterior computations are usually dominated by the theoretical predictions $\bomu(\both)$. We therefore recommend storing all computed values of $\bomu(\both)$ when computing the blinded posterior. Deblinding is then achieved with the following numerically lightweight post-processing of the blinded posterior.

\subsection{Deblinding by posterior post-processing}
During the blinded analysis, a blinded posterior $\breve{\calP}(\both|\breve{\fatx},\breve{\boSig})$ was computed. The aim is now to compute the unblinded posterior $\calP(\both|\fatx,\boSig)$ whilst minimizing computational overload. This is achieved by multiplying the posterior with deblinding weights, which corresponds to importance sampling.

We relate the blinded and deblinded posterior by
\begin{equation}
 \calP(\both|\fatx,\boSig) = w(\both)\breve{\calP}(\both|\breve{\fatx},\breve{\boSig}).
 \label{debmul}
\end{equation}
The blinded posterior will have been computed at $N$ discrete points $\both_i$, with $i \in [1,N]$. If the posterior was computed on a grid, then the $\both_i$ are regularly spaced; if the posterior was sampled with a Monte Carlo Markov Chain (MCMC) technique, then the index $i$ enumerates the samples of the chain. The aim is now to avoid that a new chain must be run after deblinding, as this may be extremely costly.

We therefore use that the general MCMC sampler will have produced a chain which will equilibrate to the \emph{blinded} posterior, but each sample will additionally have a weight which depends on the exact algorithm used. This weight can be multiplied with deblinding weights, in order to produce chains which equilibrate to the deblinded posterior. Each sample of the deblinded chain will then contribute with more or less weight to the deblinded posterior.

\begin{figure}
\includegraphics[width=0.45\textwidth]{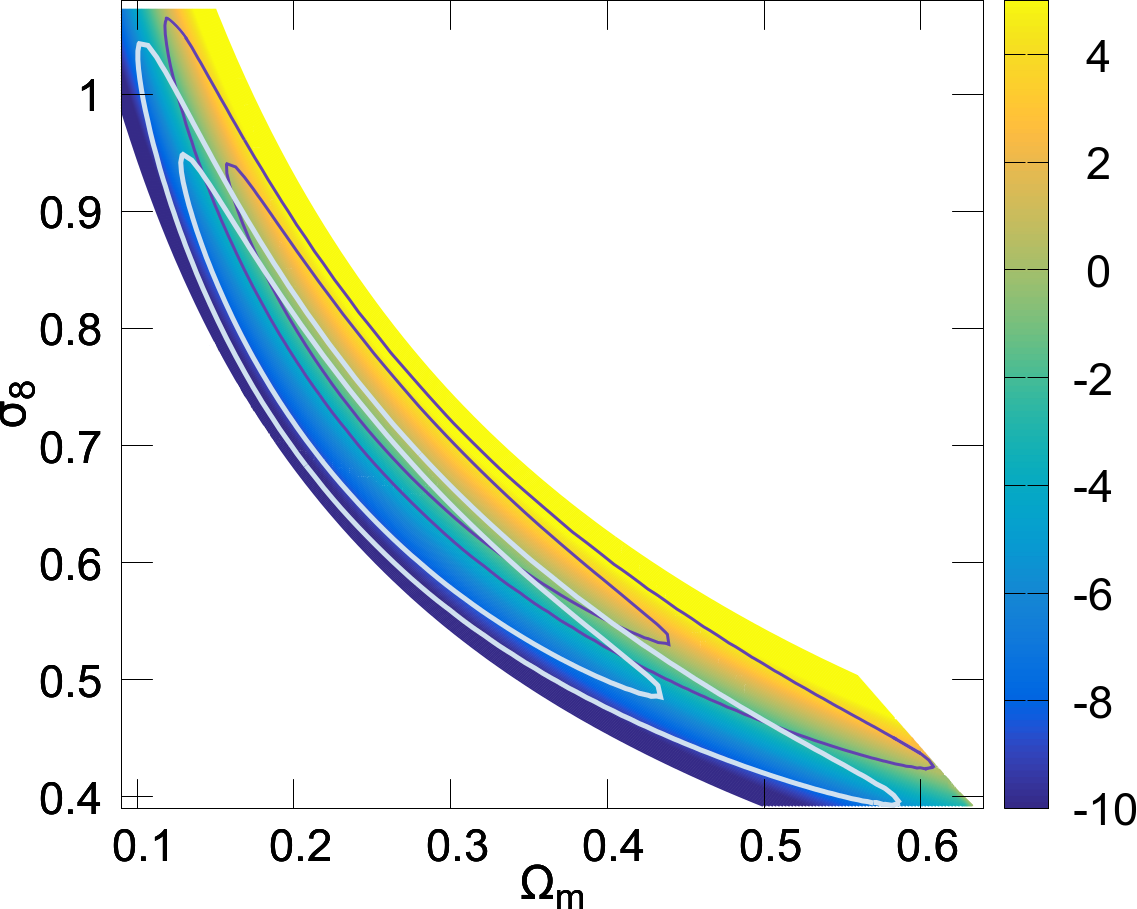} 
\caption{Illustration of deblinding: the colourbar refers to the logarithm of the deblinding weights from Eq.~(\ref{deblindingweights}). Multiplication of the bright-blue blinded posterior with the deblinding-weights results in the deblinded dark-purple posterior. Where the blinded and true posterior overlap, the de-blinding weights take values around units (zero on the plotted log-scale). Positive weights (yellow) indicate an increase of posterior probability during deblinding. Negative weights (blue and purple) indicate downweighted parameter probabilities during deblinding.}
\label{Deblinded}
\end{figure}

For all samples, the deblinding weights are
\begin{equation}
    w(\both_i) = \frac{\calP(\both_i|{\fatx},{\boSig})}{\breve{\calP}(\both_i|\breve{\fatx},\breve{\boSig})}.
    \label{deblindingweights}
\end{equation}
As the means $\bomu(\both_i)$ were stored, the weights $w(\both_i)$ are quickly evaluated. For a grid-based posterior computation, the blinded posterior is directly multiplied with the weights, according to Eq.~(\ref{debmul}). For the MCMC sampled case, the former  weights of each sample are multiplied by the deblinding weights. Deblinding by posterior weighting is depicted in Fig.~\ref{Deblinded}.

We caution that MCMC convergence after deblinding should be carefully assessed. It may be advisable to hide one unblinded analysis amongst other blinds, to enforce a high sample density in the region of importance.

\section{Discussion}
\label{sec:conclusions}
This publication established a numerically lightweight blinding and deblinding algorithm which ties in with astronomy's special circumstance of having often unique, irrepeatable, and easily recognizable data sets.

It is often said blinding avoids iterative or subconscious biasing of an analysis in order to fall in line with the status quo of a field. The positive flip-side of this view is that
blinding eases the possibility to convincingly disprove a status quo, and thereby avoid stagnation of the field. Blinding also motivates increased model-independent testing of the analysis, thereby strengthening the understanding of the astronomical data prior to inferring physics. Blinding has therefore only positive aspects, if numerically cheap, as is the case for the here presented strategy.

Our blinding strategy allows limited control over the inference to be assigned to external researchers, thereby addressing potential concerns about the details of distilling science-ready data out of astronomical raw data. In total, the presented strategy enables up to three-stage blinding, where especially the covariance-blinder has --up to parameter degeneracies-- close to perfect control over determining where the blinded posterior shall peak. 

Once sampled or computed on a grid, the posterior can be deblinded by multiplying with deblinding weights, thereby revealing the parameters actually preferred by the data. 

\appendix
\section{Essentials of cosmic shear}
\label{app:shear}
Cosmic shear is a cosmological observation technique for which we showcased the blinding algorithm. Cosmic shear measures the shearing of galaxy images on the sky. This effect arises from general relativity, according to which light follows null-geodesics which adapt to the presence of matter. This leads to light being deflected by massive objects. 

As the cosmic matter fields are perturbed, the deflection of light traversing them imprints  similar perturbation patterns on galaxy images. Cosmic shear measures these distortions over significant fractions of the sky, and computes two correlation functions from it. In this article, these two correlation functions are denoted as $\xi_+$ and $\xi_-$, and are measured as a function of angular seperation $\vartheta$ expressed in arcminutes. 

Both $\xi_+(\vartheta)$ and $\xi_-(\vartheta)$ are simultaneously caused by cosmic shear -- the presented data set in Fig.~\ref{UnblindedBestFit} therefore includes the upper ($\xi_+$) as well as the lower panels ($\xi_-$).

The triangular arrangement of the data set results from having partitioned all observed galaxies into `bins', where each bin is identified by its mean redshift. Approximately, the galaxies closest to us are assigned to redshift bin `0' (see plot labels), and the galaxies furthest from us are assigned to bin `3'. Of all bins, the auto-correlation and cross-correlation functions are measured, which leads to the labels `0 0', `0 1'..., `3 3' in the subpanels of the triangular plots. 

In total, all subpanels of Fig.~\ref{UnblindedBestFit} display one joint 130-dimensional data set. For further detail we refer the interested reader to \citet{KiDS450} and references therein.

\section{Implications for KiDS-450}
\label{Kids_specific}
This paper developed a general blinding technique and illustrated it on the KiDS-450 data which were first analyzed in \citet{KiDS450}. This appendix embeds our findings in the larger context of cosmic shear research.

Cosmic shear is highly sensitive to the dark matter density $\Omega_{\rm m}$ and the power spectrum amplitude $\sigma_8$ via the combination $S_8 = \sigma_8\sqrt{\Omega_{\rm m}/0.3}$. A mild tension between cosmic shear constraints for $S_8$ and Planck constraints on $S_8$ \citep{Planck2018} has persisted for multiple years now, with cosmic shear returning lower values of $S_8$ than Planck. Our shift of the original KiDS-450 posterior (yellow in  Fig.~\ref{KiDS-450Posterior}) towards higher $S_8$ (red in Fig.~\ref{KiDS-450Posterior}) should therefore be put into context.

Fig.~\ref{SmapeCovs} illustrates that SMAPE errors of at most 0.12 (defined in Eq.~(\ref{eq::smape}) and similar to percentual deviations) for covariance matrix elements suffice to allow shifts of the KiDS posterior into Planck compatible regions. As cosmic shear covariance matrices are either analytical approximations or numerical estimates, they will indeed be biased to a certain degree, but currently no evidence exists that the very specific bias required for the posterior shift affects the KiDS-450 covariance matrix.

Nonetheless, it is surprising that all data points of the highest redshift bin in KiDS-450 light up \emph{consistently} in Fig.~\ref{SmapeCovs}. Usually, the method here presented will affect all data points to a low degree without any preference of physically meaningful subgroups in the data.

It is thus unclear why Fig.~\ref{SmapeCovs} consistently impacts the highest redshifts. The safest interpretation of Fig.~\ref{SmapeCovs} is that the $S_8$ tension correlates with the total uncertainty at high redshifts -- whether this correlation implies a causal connection is not known, but it illustrates that an agnostic route towards understanding the origin of the tension between Planck and cosmic shear has to include a detailed understanding of cosmic shear uncertainties at high redshifts.

\section{Acknowledgements}
ES is supported by Leiden's Oort-Fellowship programme and thanks Hendrik Hildebrandt, Catherine Heymans, Koen Kuijken and Henk Hoekstra. This research is based on data products from observations made with ESO Telescopes at the La Silla Paranal Observatory under programme IDs 177.A-3016, 177.A-3017 and 177.A-3018.

\bibliographystyle{mn2e}
\bibliography{TDist}

\label{lastpage} 
\bsp 
\end{document}